\documentstyle[12pt]{article}

\newcommand{\be}{\begin{equation}}
\newcommand{\ee}{\end{equation}}
\newcommand{\bea}{\begin{array}}
\newcommand{\ea}{\end{array}}
\newcommand{\beqa}{\begin{eqnarray}}
\newcommand{\eeqa}{\end{eqnarray}}
\newcommand{\bean}{\begin{eqnarray*}}
\newcommand{\eean}{\end{eqnarray*}}
\newcommand{\eqn}[1]{(\ref{#1})}

\newcommand{\del}{\partial}

\def\up#1{\leavevmode \raise.16ex\hbox{#1}}

\def\sqr#1#2{{\vcenter{\vbox{\hrule height.#2pt
        \hbox{\vrule width.#2pt height#1pt \kern#1pt
          \vrule width.#2pt}
        \hrule height.#2pt}}}}
\def\square{\mathop\sqr68}

\newcommand{\gapproxeq}{\lower .7ex\hbox{$\;\stackrel{\textstyle >}{\sim}\;$}}
\newcommand{\lapproxeq}{\lower .7ex\hbox{$\;\stackrel{\textstyle <}{\sim}\;$}}

\newcounter{appendice}

\def\thebibliography#1{{\bf REFERENCES\markboth
 {REFERENCES}{REFERENCES}}\list
 {[\arabic{enumi}]}{\settowidth\labelwidth{[#1]}\leftmargin\labelwidth
 \advance\leftmargin\labelsep
 \usecounter{enumi}}
 \def\newblock{\hskip .11em plus .33em minus -.07em}
 \sloppy
 \sfcode`\.=1000\relax}

\begin{document}

\hfill November 19, 1994

\hfill UR-1395; ER-40685-843

\setcounter{footnote}{1}

\begin{center}
\vskip 0.5cm

{\large{\bf  Correlation Functions of a Conformal Field Theory \\
in Three Dimensions}}

\vskip 0.8cm
{\bf S.~Guruswamy}\footnote{\scriptsize guruswamy@urhep.pas.rochester.edu} \\
{\it Department of Physics, University of Rochester\\
Rochester, NY 14627 USA}\\
\vskip 0.4cm
{\bf P.~Vitale}\footnote{\scriptsize vitale@axpna1.na.infn.it}\\
{\it Dipartimento di Scienze Fisiche,
Universit\`a di Napoli\\ and I.N.F.N  Sez. di Napoli\\
Mostra d'Oltremare, Pad.19 - 80125
Napoli, ITALY}
\vskip 0.2cm
\end{center}

\vskip 0.8cm

\begin{abstract}
We derive explicit forms of the two--point correlation functions
of the $O(N)$ non-linear sigma model at the critical point, in the large $N$
limit, on various three dimensional manifolds of constant curvature. The
two--point correlation function, $G(x, y)$, is the only $n$-point correlation
function which survives in this limit. We analyze the short distance and long
distance behaviour of $G(x, y)$.  It is shown that $G(x, y)$ decays
exponentially with the Riemannian distance on the spaces $R^2 \times S^1,~S^1
\times S^1 \times R, ~S^2 \times R,~H^2 \times R$. The decay on $R^3$ is of
course a power law. We show that the scale for the correlation length is given
by the geometry of the space and therefore the long distance behaviour of the
critical correlation function is not necessarily a power law even though the
manifold is of infinite extent in all directions; this is the case of the
hyperbolic space where the radius of curvature plays the role of a scale
parameter. We also verify that the scalar field in this theory is a primary
field with weight $\delta=-{1 \over 2}$; we illustrate this using the
example of the manifold $S^2 \times R$ whose metric is conformally equivalent
to that of $R^3-\{0\}$ up to a reparametrization.
\end{abstract}
\newpage
\noindent{\bf 1. Introduction}

In a previous paper \cite{3dcft} we study the $O(N)$ non-linear sigma
model in three
 dimensions, on manifolds of constant (positive, negative and zero)
curvature, of the type $\Sigma \times R$ or $\Sigma \times S^1$, where
$\Sigma$ is a two dimensional surface.
We study the model in the $1/N$ expansion, and show that it is a conformally
invariant theory in the lowest order, at the ultra--violet stable fixed point.
We also find
the critical values of the physical mass, the vacuum expectation of the field
(spontaneous magnetization) and the free energy.

In this paper we address the problem of the two--point correlation function
(Green's function) of the theory, for the manifolds mentioned above.

We give here a brief summary of the
$O(N)$ non-linear sigma model in the
large $N$ limit on three dimensional curved spaces. For a detailed description
we refer the reader to \cite{3dcft}.

The regularized euclidean
partition function of the $O(N)$ non-linear sigma model in three
dimensions, in the
 presence of a background metric, $g_{\mu \nu}(x)$,
 can be written as,
\be
{\cal Z}[g, \Lambda, {\lambda(\Lambda)}]=\int {\cal D}_{\Lambda}[\phi]
{\cal D}_{\Lambda}[\sigma]
{\rm exp} \Bigl\{ {-\int {d^3x
{\sqrt g}\Bigl [{1\over 2}{\phi^i}(-{\square}_g+{\sigma}) {\phi_i}
-{{\sigma}\Lambda \over 2{\lambda(\Lambda)}}\Bigr ]}}\Bigr\} \label{partfun}
\ee

\noindent where $i=1,2,\cdots,N$; $\lambda$ is the coupling constant
and $\Lambda$ is the ultraviolet cut-off introduced to regularize the theory;
 $~{\cal D}_{\Lambda}[\phi]
=\prod\limits_{|k|<{\Lambda}} d\phi (k)~$
and similarly $~{\cal D}_{\Lambda}[\sigma]~$.

\noindent $\square_g~$ is the conformal
laplacian: $-\square_g = -\Delta_g + \xi {\cal R}$, where ${\cal R}$
denotes the Ricci scalar
and $\xi= \frac{d-2}{4(d-1)}$; $d$  is the dimension of the manifold.
The constraint on the $\phi$ fields, $\phi^i(x) \phi_i(x)=1$,
has been implemented by a
Lagrange multiplier, in the form of an auxiliary field $\sigma$ (the
canonical dimension of $\sigma$ in mass units is $[\sigma]=2$).

Under the conformal transformation
of the metric,
$~g_{\mu \nu}(x)\rightarrow e^{2f(x)} g_{\mu \nu}(x)~$ with
$\phi(x)\rightarrow e^{(2-d)f(x)} \phi(x)$ and
$\sigma(x) \rightarrow e^{-2f(x)} \sigma(x)$ only the part of
the classical action which is quadratic in $\phi^i$ is
conformally invariant, but the quantum theory has a non-trivial
fixed point at which ${\cal Z}$ is conformally invariant.

In \cite{3dcft} the model is studied in the leading order of
the $1/N$ expansion.
For  this purpose, we redefine
$(N-1) \lambda(\Lambda)$ as $\lambda(\Lambda)$ (kept
fixed as $N\rightarrow \infty$); also, we rescale the $\phi$
field to
${\sqrt {N-1}}~\phi$, we integrate out
the first $N-1$ components of the $\phi$ field (which is always possible on
spaces of constant curvature) and the partition function is finally
rewritten as
\beqa
{\cal Z}[g,\Lambda,\lambda(\Lambda)] &=&
\int {\cal D}_{\Lambda}[\phi_N]
{\cal D}_{\Lambda}[\sigma]~
{\rm exp}\Bigl\{ -{(N-1)\over 2}
{}~\Bigl [{\rm Tr~Log}_{\Lambda} (-{\square}_g+
{\sigma}) \nonumber \\
& +& \int d^3x {\sqrt g}
[{\phi_N}(-{\square}_g+{\sigma}) {\phi_N}
-{\Lambda \over {\lambda(\Lambda)}}{\sigma(x)}]\Bigr ] \Bigr\}
\label{partfun1}.
\eeqa
The above expression is evaluated in the large $N$ limit
at the uniform saddle point:
\bean
\langle \sigma\rangle & = & m^2 \\
\langle \phi_N \rangle & = & b,
\eean
where $m^2$ and $b$ are constants representing respectively the physical
mass and the spontaneous magnetization.

They are the solutions to the following
`gap equations', which are obtained by extremizing
the action with respect to
$\phi_{N}(x)$ keeping $\sigma(x)$ fixed and vice--versa:
\beqa
(-{\square}_g+m^2) b&=&0 \nonumber \\
{\Lambda \over {\lambda(\Lambda)}}
- G_{\Lambda}(x, x; m^2,g)&=& b^{2}; \label{gap}
\eeqa

\noindent here $G(x,y;m^2,g)$ is the
two--point Green's function of the $\phi$ fields at the saddle point,
defined as
\be
(-{\square}_g + m^2) G(x, y, m^2, g) = \frac{1}{\sqrt{g}} \delta(x,y)
\label{difgre}
\ee
where $\sqrt{g}$ denotes the square root of the determinant of the metric;
also, the Green's function is formally represented as
\be
G(x, y, m^2, g) = \langle x|(-{\square}_g + m^2)^{-1}|y \rangle
\label{green}
\ee
where $|x\rangle$, $|y\rangle$ are position eigenstates.

We will be interested in solving \eqn{difgre}, for various geometries
at the non-trivial critical point $\lambda = \lambda_c$. The value of
$m$ we would be using would be the critical value given by the gap
equations at $\lambda_c$. For a derivation of the critical values of
$m$ (and $b$) we refer to \cite{3dcft}.
Here, we merely summarize the results from \cite{3dcft} as we will be using the
critical value of $m$, $m_c$, for our calculations.
\vskip .4cm
$$
\vbox
{{\def\entry#1:#2:#3:{\strut\quad#1\quad&\quad#2\quad&\quad#3\quad\cr}
\offinterlineskip\tabskip=0pt \halign{%
\vrule\quad\hfill#\hfil\quad\vrule&\quad\hfill#\hfil\quad\vrule&
\quad\hfill#\hfil\quad\vrule\cr
\noalign{\hrule}
\vphantom{\vrule height 2pt}&&\cr\noalign{\hrule}
\entry \it Manifold:\it $m_c$:
\it $b_c$:
\vphantom{\vrule height 2pt}&&\cr\noalign{\hrule}
\vphantom{\vrule height 2pt}&&\cr\noalign{\hrule}
\entry $R^3$: zero: zero:
\entry $R^2\times S_{\beta}^1$: $\frac{2 \log}{\beta}  (\frac{1 + \sqrt 5}
{2})$: zero:
\entry $S^1\times S^1\times R$: $\not= 0$: zero:
\entry $S^2\times R$: zero: zero:
\entry $H_{\rho}^2\times R$: ${1/2\rho}$: $\not= 0$:
\vphantom{\vrule height 2pt}&&\cr\noalign{\hrule}}}}
$$
\vskip .4cm

Solutions to \eqn{difgre} can be written in terms of the heat kernel of the
operator $-\square_g +m^2$, which we will denote by $h(t; x,y,m^2,g)$:
\be
G(x,y,m^2,g)= \int_0^{\infty} dt~ h(t;x,y,m^2,g) \label{greker}
\ee
where $h(t;x,y,m^2,g)=\langle x| e^{-(-{\square_g}+m^2)t} |y \rangle$.
The heat kernel is determined by the heat equation
$$
(-{\square}_g + m^2) h (t;x,y,m^2,g)=-\frac{\del}{\del t}h (t;x,y,m^2,g)
$$
with the boundary condition
$$
h (0;x,y,m^2,g)=\frac{1}{\sqrt{g}} \delta(x,y).
$$
As well known this equation is solved by
\be
h (t;x,y,m^2,g)= \sum_n e^{-\lambda_n t} \psi^*_n(x) \psi_n(y) \label{heat}
\ee
where $\lambda_n$ are the eigenvalues of $-\square_g + m^2$ and $\psi_n$ are
the eigenstates, with $\psi_n (x) = \langle x|\psi_n \rangle$; the sum is
understood to
take into account the multiplicity of the eigenvalues.
When the spectrum is continuous the sum is replaced by an integral and the
multiplicity by the density of states.
In this paper we find the heat kernel \eqn{heat}, and consequently the
Green's function \eqn{greker}, on various manifolds of constant curvature.

We note that, to the leading order in the $1 \over N$ expansion of the
generating functional, only the two--point correlation function survives,
the higher $n$-point correlation functions being subleading in the expansion
parameter $\frac{1}{N}$. This can be seen by coupling a source current $J$
to the field $\phi_N$ in the action in \eqn{partfun1}: it is easy to check
that the $n$--point correlation function of the field $\phi_N$, $\langle
\phi_N(x_1) \phi_N(x_2)\cdots \phi_N(x_n) \rangle$ is of order $1 \over
{\sqrt N}$ with respect to the $(n-1)$--point correlation function.
Therefore it is sufficient to study the one--point (spontaneous
magnetization) and two--point correlation functions for the large $N$
theory.

\vskip .5cm
\noindent{\bf 2. Manifolds with zero curvature}

We consider the following examples of flat spaces: $R^3$,
$R^2\times S^1$, $S^1\times S^1\times R$.
The first example ($R^3$), being a well known one, is sketched for future
comparisons. For $R^2\times S^1$, some results are also known (\cite{rosen}).
the Ricci scalar being zero.

\vskip .5 cm
\newpage
\noindent {\it i) The euclidean space $R^3$}

The eigenvalues of the Laplacian $-\Delta_{R^3}$ are given by $k^2$,
where $k$ takes values on the real line, so that the
heat kernel of $~-\Delta_{R^3} + m^2~$is
\beqa
h_{R^3}(t; x,y,m^2,g) &=&\int_{-\infty}^{\infty} {d^3k \over (2 \pi)^3}
e^{-(k^2+m^2)t} e^{i\bar k (\bar x -\bar y)}\nonumber\\
&=& \frac{e^{-\frac{|\bar x -\bar y|^2}{4t}}}{(4\pi t)^{\frac{3}{2}}}
e^{-m^2t}.
\eeqa
The two--point Green's function is then:
\beqa
G_{R^3}(x,y,m^2,g)&=& \int_0^{\infty} dt \frac{e^{-\frac{|\bar x -\bar y|^2}
{4t}}}{(4\pi t)^{\frac{3}{2}}}   e^{-m^2t} \nonumber\\
&=&{\frac{1}{4 \pi}}\frac{e^{-m |\bar x -\bar y|}}{|\bar x - \bar y|}.
\label{greR3}
\eeqa
At the critical point, $m=0$, namely the correlation length (inverse of the
square root of the smallest eigenvalue of $~-\square_g+m^2$) diverges, and we
recover the expected result that at the phase transition the two--point
correlation function has a power law behaviour, $~|\bar x- \bar y|^{\alpha}~$,
with $\alpha =-1$ and there is long range order in the system.

\vskip .5 cm

\noindent {\it ii) $R^2 \times S^1$}

For this and other cases where the manifold $M$ is a product of two, say
$M=A\times B$, it is convenient to recall that the heat kernel of an
operator on $M$ is expressible as the product of heat kernels on the spaces
$A$ and $B$. From \eqn{heat} we have indeed
$$
h (t;x,y,m^2,g)= \sum_{n, k} e^{-(a_n + b_k + m^2) t} \psi^*_{n,k}(x)
\psi_{n,k}(y)
$$
being $\square_g=\square_g^A+\square_g^B$ and $a_n$, $b_k$ the eigenvalues of
$\square_g^A$ and $\square_g^B$ respectively. Moreover the eigenvectors of
$\square_g$ can be written as tensor products of eigenvectors of
$\square_g^A$ and $\square_g^B$ ($|\psi_{n,k}\rangle = |\psi_n^A\rangle \otimes
|\psi_k^B \rangle$), and the position eigenstates on $M$ as tensor products of
position eigenstates on the two spaces ($|x\rangle = |x_A\rangle \otimes
|x_B \rangle$). We have then
\beqa
h (t;x,y,m^2,g)&=& \sum_{n, k} e^{-(a_n + b_k + m^2) t} \psi^{A*}_n(x_A)
\psi^A_n(y_A) \psi^{B*}_k(x_B) \psi^B_k(y_B)\nonumber\\
&=& h_A(t;x_A,y_A,g_A) h_B (t;x_B,y_B,g_B) e^{-m^2 t}. \label{heatprod}
\eeqa
In the case of $R^2 \times S^1$ \eqn{heatprod} becomes
\be
h_{R^2\times S^1} (t;x,y,m^2,g)=h_{R^2}(t;\bar x,\bar y,g_{R^2})
h_{S^1} (t;\theta,\theta^{\prime},g_{S^1}) e^{-m^2 t}
\ee
where $\bar x$, $\bar y$, are coordinates on $R^2$, $\theta$,
$\theta^{\prime}$ are
coordinates on $S^1$. The critical value of $m$ in this case is
non-zero: in \cite{sach} $m$ was
found to be $\frac{2}{\beta} \log~(\frac{1+\sqrt{5}}{2})$, where $\beta$ is
the radius of the circle (see also
\cite{3dcft}).
The heat kernels for $~-\Delta_{R^2}~$ and $~-\Delta_{S^1}$ are respectively
\beqa
 h_{R^2}(t;\bar x,\bar y,g_{R^2}) &=& \int_{-\infty}^{\infty} {d^2k \over
(2 \pi)^2}
e^{-(k^2+m^2)t} e^{i\bar k (\bar x -\bar y)}=
\frac{e^{-\frac{|\bar x -
\bar y|^2}{4t}}}{4 \pi t} \nonumber\\
h_{S^1} (t;\theta,\theta^{\prime},g_{S^1}) &=& {1 \over \beta}
\sum_{-\infty}^{\infty} e^{-\frac{4 \pi^2
n^2}{\beta^2}t} \psi ^*_n(\theta)\psi_n(\theta^{\prime});\label{heatS}
\eeqa
where $k^2$ are the eigenvalues for $~-\Delta_{R^2}~$,
$\omega_n=\frac{4 \pi^2 n^2}{\beta^2}~$ and
$~{\psi}_{n}(\theta)=\frac{e^{in\theta}}{\sqrt{2\pi}}~$ are respectively
the eigenvalues and the eigenfunctions for $~-\Delta_{S^1}$ .
Substituting in \eqn{greker} we obtain
\be
G(x,y)= {1 \over \beta}
\int_0^{\infty} dt \frac{e^{-\frac{|\bar x -
\bar y|^2}{4t}-m^2t}}{4 \pi t}  \sum_{-\infty}^{\infty} e^{-\frac{4\pi^2
n^2}{\beta^2}t} \psi^*_n(\theta)\psi_n(\theta^{\prime}).
\ee
Long range correlations, if any, could occur only along the $R^2$ direction,
therefore we may assume the angular co-ordinates of $x$ and $y$
to be the same, which
will imply $\psi^*_n(\theta)\psi_n(\theta^{\prime})|_{\theta=\theta^{\prime}}
={1 \over 2 \pi}$.
We use the Poisson sum formula
\be
{1 \over 2 \pi }
\sum_{-\infty}^{\infty} e^{-\frac{4\pi^2 n^2}{\beta^2}t}
= \frac{\beta}{(4 \pi t)^{\frac{1}{2}}}+
\frac{2\beta}{(4\pi t)^{\frac{1}{2}}}
\sum_1^{\infty} e^{-\frac{n^2 \beta^2}{4t}} \label{poisson}
\ee
to be able to do the integral in $t$ first. The integral over $t$ can be
performed by using the following standard result \cite{prud}
\be
\int_0^\infty dt\quad t^{\nu -1} e^{-({\sigma \over t}+{\gamma t})}
= 2 {\sigma \overwithdelims () \gamma}^{\nu \over 2}
{\cal K}_{\nu} (2 {\sqrt {\sigma\gamma}}) ;\quad {\rm Re}{\sigma} > 0,
{\rm Re}{\gamma} >0, \label{besselint}
\ee
where ${\cal K}_{\nu}$ is
the MacDonald's function.
We find
\be
G_{R^2\times S^1}(x,y,m^2,g)=\frac{1}{4 \pi} \frac{e^{-m|\bar x -\bar
y|}}{|\bar x -\bar y|}+ \frac{1}{2 \pi} \sum_1^{\infty}
\frac{e^{-m \sqrt{|\bar x -\bar y|^2 +n^2 \beta^2}}}{(|\bar x -\bar y|^2 +
n^2\beta^2)^{\frac{1}{2}}} \label{gre2}
\ee
where we used
\be
{\cal K}_{1 \over 2}(x)={\pi \overwithdelims () 2 x}^{1 \over 2} e^{-x}.
\label{besselk}
\ee
\noindent In the limit ${|\bar x -\bar y|} \rightarrow \infty$, $\beta$
being finite, the sum in \eqn{gre2} can be approximated by an integral
which can be evaluated using the standard result \cite{prud}:
\be
\int_{|\bar x- \bar y|}^\infty du
\frac{e^{-m u}}{(u^2-|\bar x -\bar y|^2)^{\frac{1}{2}}}=
{\cal K}_0(m |\bar x- \bar y|).\label{approxsum}
\ee
Moreover, using the asymptotic expression of the MacDonald's function for large
$|\bar x- \bar y|$
\be
{\cal K}_{\nu} (z) ~~\lower 6pt \hbox{$\buildrel \sim \over
{\scriptscriptstyle z \rightarrow \infty}$}~~
 \sqrt{\frac{\pi}{2z}} ~e^{-z}, \label{infK}
\ee
the Green's function takes the following form when $|\bar x-\bar y|
\rightarrow \infty$:
\be
G(x,y)~~\lower 6pt \hbox{$\buildrel \longrightarrow \over
{\scriptscriptstyle
{|\bar x - \bar y |\rightarrow
\infty}}$}~~ \frac{1}{4 \pi} \frac{e^{-m|\bar x -\bar
y|}}{|\bar x -\bar y|}+ \frac{1}{\sqrt{8 \pi}~ \beta}
\frac{e^{-m|\bar x -\bar y|}}{\sqrt{m|\bar x -\bar y|}}
\ee
We see therefore that the correlation function decays exponentially for
large $|\bar x-\bar y|$.
The correlation length is in this case finite, due to the finite size
of the manifold in the $S^1$ direction (we will see that a finite
correlation length at criticality is not always connected to compactness
of the manifold in some directions).

For small values of $|\bar x - \bar y |$, we should recover the result on
 $R^3$, at criticality.
$$
G(x,y) ~~\lower 6pt \hbox{$\buildrel \longrightarrow \over
{\scriptscriptstyle
{|\bar x - \bar y |\rightarrow
0}}$}~~\frac{1}{4 \pi}
\Bigl [\frac{1}{|\bar x -\bar y|} -m \Bigr ]+
{1 \over 2 \pi} \sum_{n=1}^\infty {e^{-m \beta n} \over n \beta}.
$$
This on simplification reduces to
$$
G(x,y)~~\lower 6pt \hbox{$\buildrel \longrightarrow \over
{\scriptscriptstyle
{|\bar x - \bar y |\rightarrow
\infty}}$}~~
\frac{1}{4 \pi |\bar x -\bar y|} +\log2\sinh{m \beta \over 2};
$$
but,~~  $\log2\sinh{m \beta \over 2}=0$ for the critical value of $m$.
Therefore the short distance behaviour of the Green's function on $R^2 \times
 S^1$ is
\be
G(x,y)~~\lower 6pt \hbox{$\buildrel \longrightarrow \over
{\scriptscriptstyle
{|\bar x - \bar y |\rightarrow
0}}$}~~
\frac{1}{4 \pi|\bar x -\bar y|}~
\ee
which is the same as in $R^3$, where the Green's function at criticality is
given by \eqn{greR3} with $m=0$.

\vspace*{5mm}

\noindent {\it iii) $S^1 \times S^1 \times R$}

For simplicity, we consider the circles to have the same radius, $\rho$.
Let the coordinates on $S^1
\times S^1$ be denoted by $(\theta_1, \theta_2)$
and the coordinate on $R$ by $x_0$.
The Ricci scalar is zero on this space so that
$-{\square}_g=-\Delta_{S^1 \times S^1}-\Delta_R$. From \eqn{heatprod} we
have
\be
h_{S^1\times S^1\times R} (t;x,y,m^2,g)=h_{S^1}(t;\theta_1, \theta_1^{\prime})
h_{S^1}(t;\theta_2, \theta_2^{\prime})
h_{R} (t;x_0,y_0) e^{-m^2 t} \label{heatSSR}
\ee
The heat kernel of $-\Delta_{S^1}$ has been given in \eqn{heatS} while the heat
kernel of $-\Delta_{R}$ is
\be
h_{R}(t;x_0,y_0) = \int_{-\infty}^{\infty} {dk \over
2 \pi}
e^{-(k^2+m^2)t} e^{i k | x_0 - y_0|}=\frac{e^{-\frac{( x_0 -
y_0)^2}{4t}}}{(4 \pi t)^{\frac{1}{2}}} \label{heatR}
\ee
As in the previous case,
we fix the angular separation between the points $x$ and $y$ to be zero
($\theta_1=\theta_1^{\prime}, \theta_2=\theta_2^{\prime}$),
that is, we are looking for the behaviour of the two--point correlation
function
along the $R$ direction. Substituting \eqn{heatSSR} in \eqn{greker} we have

\be
G(x,y)= \frac{1}{4 \pi^2 \rho^2}\int_{0}^{\infty} dt
\frac {e^{-\frac{|x_0-y_0|^2}{4t}}}
{\sqrt{4 \pi t}}
\sum_{p,q=-\infty}^{\infty}
 e^{-{4 \pi^2 \over \rho^2} (p^2+q^2)t} e^{-m^2 t}.
\ee
On using the Poisson sum formula
\eqn{poisson}, the two--point correlation function is
\be
G(x,y)=\sum_{p,q} \int_0^\infty {dt \over (4 \pi)^{3 \over 2}} \quad
t^{-{3 \over 2}} e^{-m^2 t} e^{-(p^2+q^2) {\rho^2 \over 4 t}}
e^{-{|x_0-y_0|^2 \over 4t}}
\ee
The modes $p,q=0$, $\{p=0,q=1\}$ and $\{p=1,q=0\}$
can be separated and the integrals
over $t$ can be performed using \eqn{besselint} and the resulting expressions
simplified using \eqn{besselk} to give
\beqa
G_{S^1\times S^1 \times R}(x,y,m^2,g)&=&{1 \over 4 \pi} {e^{-m|x_0-y_0|} \over
|x_0-y_0|}+ {1 \over \pi}
\sum_{p=1}^\infty {e^{-m{\sqrt{p^2 \rho^2+|x_0-y_0|^2}}} \over
{\sqrt{p^2 \rho^2+|x_0-y_0|^2}}}\nonumber \\
&+& {1 \over \pi}
\sum_{p,q=1}^\infty {e^{-m{\sqrt{p^2 \rho^2+q^2 \rho^2+|x_0-y_0|^2}}} \over
{\sqrt{p^2 \rho^2+q^2 \rho^2+|x_0-y_0|^2}}}. \label{gres1s1}
\eeqa
The critical value of $m$ was found in \cite{3dcft}
to be non-zero and is the solution to the gap equation
\be
-{m \rho \over 4}
+
\sum_{p=1}^\infty {e^{-m p \rho} \over
{p}}+
\sum_{p,q=1}^\infty {e^{-m \rho {\sqrt{p^2 +q^2 }}} \over
{\sqrt{p^2+q^2 }}}=0.\label{gaps1s1r}
\ee
We can now analyze the behaviour of the Green's function in the limits
$|x_0-y_0| \rightarrow \infty$ and $|x_0-y_0| \rightarrow 0$.
When $|x_0-y_0| \rightarrow \infty$ ($\rho$ being finite),
we again approximate
the sums in \eqn{gres1s1} with the corresponding integrals which can be
performed to give
\be
G(x,y)
{}~~\lower 6pt \hbox{$\buildrel \longrightarrow \over
{\scriptscriptstyle
{|x_0 - y_0 |\rightarrow
\infty}}$}~~{1 \over 4 \pi} {e^{-m|x_0-y_0|} \over |x_0-y_0|}
+\frac{1}{\sqrt{2 \pi}~ \rho} {e^{-m|x_0-y_0|} \over \sqrt{m |x_0-y_0|}}
+{1 \over 2 m \rho^2} e^{-m|x_0-y_0|}
\ee
showing that the correlation function decays exponentially and the
correlation length is finite at criticality along the $R$
direction. This again is
due to the finite size effect of the torus.

In the limit $|x_0-y_0| \rightarrow 0$,
\be
G(x,y)~~\lower 6pt \hbox{$\buildrel
\longrightarrow \over
{\scriptscriptstyle
{|x_0 - y_0 |\rightarrow
0}}$}~~{1 \over 4 \pi}\Bigl [ {1 \over |x_0-y_0|}-m \Bigr ]
+ {1 \over \pi \rho}
\sum_{p=1}^\infty {e^{-m p \rho} \over {p}}
+ {1 \over \pi \rho}
\sum_{p,q=1}^\infty {e^{-m \rho {\sqrt{p^2 +q^2 }}} \over
{\sqrt{p^2+q^2 }}}.
\ee
At the critical point, on using the gap equation \eqn{gaps1s1r},
this expression reduces to
\be
G(x,y)~~\lower 6pt \hbox{$\buildrel
\longrightarrow \over
{\scriptscriptstyle
{|x_0 - y_0 |\rightarrow
0}}$}~~{1 \over 4 \pi |x_0-y_0|}
\ee
which is the
the correct behaviour of the Green's function in
this limit as it is the same as that on $R^3$ (\eqn{greR3} with $m=0$),
when the separation between the
angular co-ordinates of the points $x$ and $y$ is zero.

\vskip .5cm

\noindent {\bf 3. A manifold with positive curvature: $S^2 \times R$}

This example is of particular interest as it provides us with a setting
to describe
what a primary field is in this three dimensional context.

We indicate with $(\theta, \phi)$
the coordinate on the sphere and with $x_0$ the coordinate on
the real line. $\rho$ is the radius of the sphere. The conformal
Laplacian on the sphere is $~-{\square}_{S^2}=-\Delta_{S^2}+{1 \over 8}
\cal R~$ where
$~{\cal R}={2 \over \rho^2}~$.
 The eigenvalues of $-{\square}_{S^2}$ are ${(l+\frac{1}{2})^2}$ with
degeneracy $(2l+1)$ where $l=0,1,2,\cdots, \infty$
and the eigenfunctions are the
spherical harmonics denoted by $Y_l^m(\theta,\phi),~ m=-l, -l+1, \cdots,l$.
The heat kernel of $-\square_{S^2}$ is then
\be
h_{S^2}(t,\bar x,\bar y)=
{1 \over \rho^2} \sum_{l=0}^\infty \sum_{m=-l}^l
 e^{-(l+{1\over 2})^2 {t \over
\rho^2}}{Y^*}_l^m(\theta,\phi)
Y_l^m(\theta^{\prime},\phi^{\prime}) \label{heatS2}
\ee
while the heat kernel of $-\Delta_R$ is given by \eqn{heatR}. The critical
value of the mass is zero \cite{3dcft} for this case.
In order to simplify the problem, let us look for the heat kernel and
hence the Green's function when the angular separation
$\theta-\theta^{\prime}$ and $\phi-\phi^{\prime}$ between the points
$x$ and $y$ is zero. Again, we can do this without loss of
generality as it is
only meaningful to look for the presence of long range correlations
along the $R$ direction.
Recalling that
\be
\sum_{m=-l}^l
{Y^*}_l^m(\theta,\phi)
Y_l^m(\theta,\phi)={2l+1 \over 4 \pi},
\ee
and substituting \eqn{heatR} and \eqn{heatS2} in \eqn{greker}, we find the
Green's function to be
\be
G(x,y)=
{1 \over 4 {\pi}^{3 \over 2}\rho^2} \int_0^\infty dt \quad
t^{-{1 \over 2}}e^{-{|x_0-y_0|^2 \over 4 t}}
\sum_{l=0}^\infty (l+{1 \over 2})
 e^{-(l+{1\over 2})^2 {t \over \rho^2}}.
\ee
We now use an extension of the Poisson sum formula to this case \cite{3dcft},
\be
{1 \over 2 \pi} \sum_{l={1 \over 2}}^\infty l e^{-{l^2 \over \rho^2} t}
={\rho^2 \over (4 \pi t)^{3 \over 2}}
\int_{-\infty}^\infty dz \quad ({z \over 2 \rho}
{\rm cosec}{z \over 2 \rho}-1) e^{-{z^2 \over 4 t}} + {\rho^2 \over 4 \pi t}
,\ee
to rewrite $G(x,y)$ as
\be
G(x,y)={ 1 \over 8 { \pi}^{3 \over 2}} \int_0^\infty dt \quad
e^{-{|x_0-y_0|^2 \over 4 t}} \Bigl [ t^{-{3 \over 2}} +
{t^{-2} \over {\sqrt{4 \pi}}}
\int_{-\infty}^\infty dz \quad ({z \over 2 \rho}
{\rm cosec}{z \over 2 \rho}-1) e^{-{z^2 \over 4 t}}  \Bigr ].
\ee
The integral over $t$ can be performed easily and the Green's function
simplifies to
\be
G(x,y)={1 \over 4 \pi} \Bigl [ {1 \over |x_0-y_0|}+
{1 \over \pi} \int_{-\infty}^\infty {dz \over {z^2+|x_0-y_0|^2}}
\quad ({z \over 2 \rho}{\rm cosec}{z \over 2 \rho}-1)
\Bigr].
\ee
The integral over $z$ can now be performed by going to the complex plane
and the final expression for the Green's function is
\be
G_{S^2 \times R}(x,y,g)={1 \over 8 \pi \rho} {\rm cosech}{|x_0-y_0| \over 2
\rho}.
\ee
When $|x_0-y_0| \rightarrow \infty$,
\be
G(x,y) ~~\lower 6pt \hbox{$\buildrel \longrightarrow \over
{\scriptscriptstyle
{| x_0 - y_0 |\rightarrow
\infty}}$}~~{1 \over 4 \pi \rho} e^{-{|x_0-y_0| \over 2 \rho}}.
\ee
There is no long range correlation even along the $R$ direction at the
critical point.
Also, when $|x_0-y_0| \rightarrow 0$,
\be
G(x,y) ~~\lower 6pt \hbox{$\buildrel \longrightarrow \over
{\scriptscriptstyle
{|x_0 - y_0 |\rightarrow
0}}$}~~{1 \over 4 \pi |x_0-y_0|}
\ee
which is the flat space limit as one should expect.

Let us now recall the definition of a primary field as given in \cite{jain},
\cite{3dcft}.
 A primary field is one
whose correlation functions transform homogeneously under
a conformal transformation:
\be
        <\phi(x_1)\cdots \phi(x_n)>_{e^{2f}g}
=e^{\delta(f(x_1)+\cdots f(x_n))}       <\phi(x_1)\cdots \phi(x_n)>_{g}.
\label{primary}
\ee
where $\delta$ is the conformal weight.
The field $\phi$ of the $O(N)$ sigma model
is an example of such a primary field. The example on $S^2 \times R$ is
particularly suitable for illustrating this fact because the metric on
this space is conformally equivalent to that on $R^3-\{0\}$ after a
reparametrization. This can be easily checked:
\noindent let the metric on
$S_\rho^2 \times R$ be denoted by $\tilde g_{\mu \nu}$
and that on $R^3-\{0\}$ by $g_{\mu \nu}$.
The line element on $S_\rho^2 \times R$ is,
$$ds_{S^2 \times R}^2=\rho^2 (du^2+d\Omega^2)$$ where, $u$ is the co-ordinate
 on $R$ and
$d\Omega$ is the solid angle.
The line element on $R^3-\{0\}$ in spherical polar co-ordinates is,
$$ds_{R^3-\{0\}}^2=dr^2+r^2 d\Omega^2.$$
On writing $r$ as $r=\rho e^u$, the line element on $R^3-\{0\}$ becomes,
$$ds_{R^3-\{0\}}^2=\rho^2 e^{2u}(du^2+ d\Omega^2).$$
We therefore see that the metrics $g_{\mu \nu}$
and $\tilde g_{\mu \nu}$ are related by a
conformal transformation,
$\tilde g_{\mu \nu}=e^{2 f} g_{\mu \nu}$ with $f(x)=-u$.
Thus the manifolds $S^2 \times R$ and $R^3-\{0\}$
are conformally equivalent.
We see that ${|x_0-y_0| \over \rho}=|u-v|$ where $u$ and $v$ are
dimensionless, real--valued quantities. In terms of the dimensionless
variables $u,~v$
\be
G_{\tilde g}(x,y)= {1 \over 8 \pi \rho} {\rm cosech}{|u-v| \over 2}.
\ee
Equation \eqn{primary} gives a general definition of a primary field. The
specific case of this, when we look at the two--point correlation function,
can be easily derived and the weights fixed.
The two--point correlation function $G_g(x,y)$ is the solution to the equation
\be
(-{\square}_g+m^2) G_g(x,y)={1 \over \sqrt g} \delta(x-y)
\ee
and $G_{\tilde g}(x,y)$ is the solution to
\be
(-{\square}_{\tilde g}+{\tilde m}^2) G_{\tilde g}(x,y)=
{1 \over \sqrt {\tilde g}} \delta(x-y).\label{transform}
\ee
Under the conformal transformation of the metric, $g_{\mu \nu}
\rightarrow {\tilde g_{\mu \nu}}=
e^{-2u} g_{\mu \nu}$, we have
$ {\sqrt {\tilde g}}=e^{-3 u} {\sqrt g}$ and
\be
(-{\square}_g+m^2) \rightarrow (-{\square}_{\tilde g}+{\tilde m}^2)
\ee
where
\be
(-{\square}_{\tilde g}+{\tilde m}^2)=e^{{5u \over 2}}(-{\square}_g+m^2)
e^{-u \over 2}.
\ee
(See reference \cite{parker} for the transformation property of the
conformal Laplacian.)
Using all this in \eqn{transform} we get the transformation property for the
Green's function:
\be
e^{-{u \over 2}}G_{\tilde g}(x,y)e^{-{v \over 2}}=G_{g}(x,y).
\ee
\be
e^{-{u \over 2}} G_{\tilde g}(x,y) e^{-{v \over 2}} =
{1 \over 4 \pi |r-r^{\prime}|}
\ee
where $r= \rho e^u$ and $r^{\prime}=\rho e^v$.
The right hand side is nothing but the Green's function on $R^3-\{0\}$
when the angular separation between two points $x$ and $y$ is zero.
This shows that the $\phi$ fields are primary fields and transform with
 weight $\delta=-{1 \over 2}$.

\vskip .5cm
\noindent{\bf 4. A manifold with negative curvature: $H^2 \times R$}

We consider a space which is a product of two non-compact manifolds, $H^2$
being the surface of a three dimensional hyperboloid, $R$ the real line.
We denote the coordinate on $R$ by $x_0$ and parametrize $H^2$ as
$$
H^2= \{ \bar x=(x_1, x_2),~ x_1 \in R,~ 0<x_2<\infty \}
$$
with line element and Laplacian given respectively by
\beqa
ds^2&=&\frac{\rho^2}{x_2^2} (dx_1^2 + dx_2^2) \nonumber \\
\Delta_{H^2}&=&\frac{x_2^2}{\rho^2}(\del_{x_1}^2 + \del_{x_2}^2)
\label{defh2}
\eeqa
 where $\rho$ is a constant positive parameter. The scalar curvature for $H^2$
(hence for the product manifold) is $ {\cal R} = -\frac{2}{\rho^2}$;
therefore $\xi{\cal R}=-\frac{1}{4\rho^2}$.
At the critical point the value of $m^2$ is $m^2_c=\frac{1}{4\rho^2}~$,
which exactly cancels with $\xi{\cal R}$. We have then:
\be
-{\square}_g + m^2_c = - \Delta_R - \Delta_{H^2} + \xi {\cal R} + m^2_c =
- \Delta_R - \Delta_{H^2}.
\ee
This implies that the heat kernel of $-{\square}_g + m^2_c$ is just the
product of the heat kernel of $- \Delta_R$, given by \eqn{heatR},
and the heat kernel \cite{davies} of $ - \Delta_{H^2}$ which is
\be
h_{H^2} (t; \bar x, \bar y, g_{H^2})=
\rho \frac {{2^{1 \over 2}}e^{-\frac{t}{4\rho^2}}}
{(4 \pi t)^{\frac{3}{2}}}\int_{\frac{d}{\rho}} ^\infty \frac{\tau
e^{-\frac{\tau^2 \rho^2}{4t}}}{\sqrt{\cosh~\tau - \cosh~\frac{d}{\rho}}}
d\tau;\label{heatH}
\ee
$d$ is the geodesic distance on $H^2$ defined as
\be
\cosh~\frac{d}{\rho}(\bar x, \bar y)= 1 + \frac{|\bar x- \bar y|^2}{2x_2 y_2}.
\ee
Substituting \eqn{heatR} and \eqn{heatH} in \eqn{greker} we get
$$
G (x, y)= \frac{\rho}{8 {\sqrt 2} \pi^2}
 \int_{\frac{d}{\rho}} ^\infty d\tau\frac{\tau}
{\sqrt{\cosh~\tau - \cosh~\frac{d}{\rho}}}
\int_0^{\infty}
dt~ t^{-2}
e^{\frac{-|x_0-y_0|^2-\tau^2 \rho^2}{4t}} e^{-\frac{t}{4\rho^2}};
$$
the integral in $t$ can be performed using \eqn{besselint} and we find

\be
G_{H^2\times R} (x, y, g)= \frac{1}{4 {\sqrt 2}~\pi^2\rho}
\int_{\frac{d}{\rho}}^\infty d\tau~ \frac{{\cal
K}_1(\frac{1}{2}\sqrt{\tau^2+\frac{|x_0-y_0|^2}{\rho^2}}) }
{\sqrt{\cosh~\tau-\cosh~\frac{d}{\rho}}}
\frac{\tau}{\sqrt{\tau^2+\frac{|x_0-y_0|^2}{\rho^2}}} \label{greHR}
\ee
where ${\cal K}_1(z)$ is a MacDonald' s function.

The large distance behaviour of the two--point correlation function can be
analyzed in the $H^2$ and $R$ directions separately.
For $d(\bar x, \bar y) \rightarrow \infty$, $|x_0 - y_0|$ remaining finite,
$\tau$ is also large (being $\tau \ge d$), so that we can approximate
the MacDonald's function by its asymptotic expression given in \eqn{infK}.
We have
$$
{\cal K}_1 (\frac{1}{2}\sqrt{\tau^2+|x_0-y_0|^2})
{}~~\lower 6pt \hbox{$\buildrel \longrightarrow \over
{\scriptscriptstyle
 \tau \rightarrow
\infty}$}~~
{\cal K}_1 (\frac{\tau}{2})
{}~~\lower 6pt \hbox{$\buildrel \longrightarrow \over
{\scriptscriptstyle
 \tau \rightarrow
\infty}$}~~
\sqrt{\frac{\pi}{\tau}}~
e^{-\frac{\tau}{2}}~;
$$
in this approximation the integral in $\tau$ in \eqn{greHR} can be performed
and we find
\be
G(x, y)
{}~~\lower 6pt \hbox{$\buildrel \longrightarrow \over
{\scriptscriptstyle
 d \rightarrow
\infty}$}~~
\frac{1}{4 \pi^{3 \over 2}\rho}~ e^{-\frac{d}{2\rho}}~ {\cal
K}_0(\frac{d}{2\rho});
\ee
on using again \eqn{infK},
we  finally obtain
\be
G (x, y)
{}~~\lower 6pt \hbox{$\buildrel \longrightarrow \over
{\scriptscriptstyle
 d \rightarrow
\infty}$}~~
\frac{1}{4 \pi \rho}~ \frac{e^{-\frac{d}{\rho}}}{\sqrt \frac{d}{\rho}}.
\ee

Let us analyze the other limit, $|x_0-y_0| \longrightarrow \infty$ keeping
$d$ fixed, in particular we choose it to be zero.
In this case, we can still use the large $z$ approximation for ${\cal
K}_{\nu} (z)$
and \eqn{greHR} becomes
\beqa
G(x, y)
&\lower 6pt \hbox{$\buildrel \longrightarrow \over
{\scriptscriptstyle
|x_0-y_0| \rightarrow
\infty}$}&
\frac{1}{8 \pi^{3 \over 2} \rho}
\int_{\scriptscriptstyle\frac{|x_0-y_0|}{\rho}}^\infty d\tau~
\frac{e^{-\frac{\tau}{2}}}
{\sqrt {\tau}\sinh~{1 \over 2} \sqrt{\tau^2-{|x_0-y_0|^2 \over \rho^2}}}
\nonumber \\
%& \lower 6pt \hbox{$\buildrel \longrightarrow \over
%{\scriptscriptstyle
%|x_0-y_0| \rightarrow
%\infty}$}&~~
&\simeq& \frac{1}{ 8\pi^2 \rho} {\cal K}_{\frac{1}{4}}^2 (\frac{1}{4\rho}
|x_0-y_0|)
\eeqa
and using \eqn{infK}
\be
G (x, y)
{}~~\lower 6pt \hbox{$\buildrel \longrightarrow \over
{\scriptscriptstyle
|x_0-y_0| \rightarrow
\infty}$}~~{\frac{1}{4 \pi}~\frac{e^{-\frac{|x_0-y_0|}{2\rho}}}
{|x_0-y_0|}}.
\ee

In both the cases analyzed  ($d \rightarrow \infty$ and $|x_0 - y_0|
\rightarrow \infty$), the large distance behaviour of the two--point
correlation function is not a
power law, that is the correlation length is not infinite, as we might
have expected from the manifold being a non-compact one in all the
directions and from the fact that there is a non-zero spontaneous
magnetization $b$ as we showed in \cite{3dcft}.
We find instead an exponential decay where the finite
correlation length is proportional to $\rho$. We conclude that the
finiteness of the correlation length, which is generally associated with the
finite size of the manifold, is more precisely connected to the presence of
a scale in the theory, in this case the scale being the radius of curvature
of the manifold.

Let us finally analyze the short distance behaviour of the two--point Green's
function \eqn{greHR}. For simplicity we consider separately the two limits
$d\rightarrow 0, ~|x_0-y_0|=0$ and $|x_0-y_0| \rightarrow 0, ~d=0$.

For $d\rightarrow 0, ~|x_0-y_0|=0$, \eqn{greHR} can be approximated as
\be
G(x, y)~~\lower 6pt \hbox{$\buildrel \sim \over
{\scriptscriptstyle
 d \rightarrow 0}$}~~
\frac{1}{8 \pi^2\rho}
\int_{\frac{d}{\rho}}^\infty d\tau~ \frac{{\cal
K}_1(\frac{1}{2}\tau) }
{\sinh~\frac{\tau}{2}}.  \label{smallgre}
\ee
The integrand is very rapidly converging to zero as $\tau$ increases. We can
split the integral in two parts and
use two different approximations for the MacDonald's function:
\be
{\cal K}_\nu(x) ~\sim~
2^{\nu-1} (\nu -1)! x^{-\nu},~~~~~\nu>0  \label{smallK}
\ee
valid in the region $x \le 1$, and  \eqn{infK} which is valid for $x>1$.
We have
\be
G(x, y)~~\lower 6pt \hbox{$\buildrel \longrightarrow \over
{\scriptscriptstyle
 d \rightarrow 0}$}~~ \frac{1}{8 \pi^2\rho} \Bigl\{
\int_{\frac{d}{\rho}}^1 d\tau~ \frac{{\cal
K}_1(\frac{1}{2}\tau) }
{\sinh~\frac{\tau}{2}} +
\int_1^\infty d\tau~ \frac{{\cal
K}_1(\frac{1}{2}\tau) }{\sinh~\frac{\tau}{2}} \Bigr\};
\ee
on substituting \eqn{smallK} in the first integral and \eqn{infK} in the
second one we find
\be
G(x, y)~~\lower 6pt \hbox{$\buildrel \longrightarrow \over
{\scriptscriptstyle
 d \rightarrow 0}$}~~ \frac{1}{2 \pi^2\rho} \Bigl\{
\Bigl (\frac{d}{\rho}\Bigr )^{-1} + {\rm
finite~terms} \Bigr \}.  \label{smagred}
\ee
Similarly, for $|x_0-y_0|\rightarrow 0, ~d=0,$ \eqn{greHR} can be approximated
as
\beqa
G(x, y)&\lower 6pt \hbox{$\buildrel \longrightarrow \over
{\scriptscriptstyle
 |x_0-y_0| \rightarrow 0}$}&
\frac{1}{8 \pi^2\rho}
\int_0^\infty d\tau~ \frac{{\cal
K}_1(\frac{1}{2}\sqrt{\tau^2+\frac{|x_0-y_0|^2}{\rho^2}}) }
{\sinh~\frac{\tau}{2}}
\frac{\tau}{\sqrt{\tau^2+\frac{|x_0-y_0|^2}{\rho^2}}} \nonumber \\
&=&\frac{1}{8 \pi^2\rho}
\int_{\frac{\scriptscriptstyle|x_0-y_0|}{\rho}}^\infty d\tau^{\prime}~
\frac{{\cal
K}_1(\frac{1}{2}\tau^{\prime}) }
{\sinh~\frac{\tau^{\prime}}{2}}.
\eeqa
where $~\tau^{\prime}=\sqrt{\tau^2+\frac{|x_0-y_0|^2}{\rho^2}}~$. This integral
is identical to \eqn{smallgre}, therefore it can be evaluated in the same
way yielding
\be
G(x, y)~~\lower 6pt \hbox{$\buildrel \longrightarrow \over
{\scriptscriptstyle
|x_0-y_0| \rightarrow 0}$}~~ \frac{1}{2 \pi^2\rho} \Bigl\{
\Bigl ( \frac{|x_0-y_0|}{\rho}\Bigr )^{-1} + {\rm
finite~terms} \Bigr \}. \label{smagreR}
\ee

Both the results \eqn{smagred} and \eqn{smagreR} confirm the expected
flat space behaviour of the two--point Green's function for short distances.

\vskip .3cm

\noindent{\bf Acknowledgements}

We thank S. G. Rajeev for suggestions and discussions.
P. Vitale thanks the Dept. of Physics of the University of Rochester
for hospitality. This work was supported in part by the US Dept. of
Energy Grant No. DE-FG02-91ER40685.

\vskip 1 cm

%
%%%%%%%%%%%%%%%%%%%%%%%%%%%%%%%%%%%%%%%%%%%%%%%%%%%%%%%%%%%%%%%
%
%      List of references
%
%%%%%%%%%%%%%%%%%%%%%%%%%%%%%%%%%%%%%%%%%%%%%%%%%%%%%%%%%%%%%%%%
%

\end{document}